\documentclass[sn-vancouver,default,iicol]{sn-jnl}

\usepackage{natbib}

\jyear{2022}%

\theoremstyle{thmstyleone}%
\usepackage{mathtools}
\usepackage{amssymb}
\usepackage{multirow}
\usepackage{amsmath}
\usepackage{braket}
\usepackage{mathrsfs}
\theoremstyle{thmstyletwo}%

\theoremstyle{thmstylethree}%

\begin{document}



\title{Evidence of free-bound transitions in warm dense matter and their impact on equation-of-state measurements}


\author*[1,2,3]{\fnm{Maximilian P.} \sur{B\"ohme}}
\email{m.boehme@hzdr.de}

\author[4]{\fnm{Luke B.} \sur{Fletcher}}

\author[5]{\fnm{Tilo} \sur{D\"oppner}}

\author[6,2]{\fnm{Dominik} \sur{Kraus}}

\author[7]{\fnm{Andrew D.} \sur{Baczewski}}

\author[8]{\fnm{Thomas~R.} \sur{Preston}}


\author[5]{\fnm{Michael J.} \sur{MacDonald}}

\author[5]{\fnm{Frank R.} \sur{Graziani}}

\author[1,2]{\fnm{Zhandos~A.} \sur{Moldabekov}}

\author[2]{\fnm{Jan} \sur{Vorberger}}

\author*[1,2]{\fnm{Tobias} \sur{Dornheim}}\email{t.dornheim@hzdr.de}

\affil[1]{\orgname{Center for Advanced Systems Understanding (CASUS)}, \orgaddress{\city{G\"orlitz}, \postcode{D-02826}, \country{Germany}}}

\affil[2]{\orgname{Helmholtz-Zentrum Dresden-Rossendorf (HZDR)}, \orgaddress{\city{Dresden}, \postcode{D-01328},  \country{Germany}}}

\affil[3]{\orgname{Technische  Universit\"at  Dresden}, \orgaddress{\city{Dresden}, \postcode{D-01062}, \country{Germany}}}

\affil[4]{\orgname{SLAC National Accelerator Laboratory}, \orgaddress{\city{Menlo Park}, \postcode{California 94309}, \country{USA}}}

\affil[5]{\orgname{Lawrence Livermore National Laboratory}, \orgaddress{\city{Livermore}, \postcode{California 94550}, \country{USA}}}

\affil[6]{\orgname{Institut f\"ur Physik, Universit\"at Rostock}, \orgaddress{\city{Rostock}, \postcode{D-18057}, \country{Germany}}}

\affil[7]{\orgname{Sandia National Laboratories}, \orgaddress{\city{Albuquerque}, \postcode{NM 87185}, \country{USA}}}

\affil[8]{\orgname{European XFEL}, \orgaddress{\city{Schenefeld}, \postcode{D-22869}, \country{Germany}}}



\abstract{ 
Warm dense matter (WDM) is now routinely created and probed in laboratories around the world, providing unprecedented insights into conditions achieved in stellar atmospheres, planetary interiors, and inertial confinement fusion experiments. However, the interpretation of these experiments is often filtered through models with systematic errors that are difficult to quantify. Due to the simultaneous presence of quantum degeneracy and thermal excitation, processes in which free electrons are de-excited into thermally unoccupied bound states transferring momentum and energy to a scattered x-ray photon become viable. Here we show that such free-bound transitions are a particular feature of WDM and vanish in the limits of cold and hot temperatures. The inclusion of these processes into the analysis of recent X-ray Thomson Scattering experiments on WDM at the National Ignition Facility and the Linac Coherent Light Source significantly improves model fits, indicating that free-bound transitions have been observed without previously being identified. This interpretation is corroborated by agreement with a recently developed model-free thermometry technique and presents an important step for precisely characterizing and understanding the complex WDM state of matter.
}

\keywords{warm dense matter, laboratory astrophysics, X-ray Thomson scattering, equation-of-state}



\maketitle

The study of matter at extreme temperatures ($T\sim10^{3}-10^8\,$K) and pressures ($P\sim1-10^4\,$Mbar) constitutes a highly active frontier at the interface of a variety of research fields including plasma physics, electronic structure, material science, and scientific computing~\cite{wdm_book,drake2018high,Hatfield_Nature_2021}. Such \emph{warm dense matter} (WDM) occurs in astrophysical objects~\cite{Bailey2015} such as giant planet interiors~\cite{Liu2019,Brygoo2021,Kraus_Science_2022} and brown dwarfs~\cite{Kritcher_Nature_2020,becker}. For terrestrial applications, WDM is of prime relevance for materials synthesis and discovery, with the recent observation of diamond formation at high pressures~\cite{Kraus2016,Kraus2017} being a case in point.
Additionally, the WDM regime must be traversed on the way to ignition~\cite{hu_ICF} in inertial confinement fusion (ICF)~\cite{Betti2016}, where recent breakthroughs~\cite{Zylstra2022} promise a potential abundance of clean energy in the future.

As a direct consequence of this remarkable interest, WDM is nowadays routinely created in large research centers around the globe, including the European XFEL in Germany~\cite{Tschentscher_2017}, SACLA in Japan~\cite{SACLA_2011}, as well as LCLS~\cite{LCLS_2016}, the OMEGA laser~\cite{OMEGA}, the Z Pulsed Power Facility~\cite{sinars2020review}, and the National Ignition Facility (NIF)~\cite{Moses_NIF,MacDonald_POP_2023} in the USA. Yet, the rigorous interpretation of WDM experiments constitutes a formidable challenge. Specifically, the extreme conditions often prevent the direct measurement even of basic parameters such as temperature and density, which have to be inferred from other observations~\cite{drake2018high}. In this situation, x-ray Thomson scattering (XRTS)~\cite{siegfried_review} has emerged as a powerful and capable tool that provides unprecedented insights into the behaviour of WDM~\cite{Gregori_PRE_2003,PhysRevE.94.053211,GarciaSaiz2008,Tilo_Nature_2023,Dornheim_T_2022,dornheim2023xray}. 
The measured XRTS intensity for a probe energy of $\omega_0$ is given by the convolution of the combined source and instrument function $R(\omega)$ with the electronic dynamic structure factor $S_{ee}(\mathbf{q},\omega)$, $I(\mathbf{q},\omega)=R(\omega)\circledast S_{ee}(\mathbf{q},\omega_0-\omega)$, where the latter, in principle, contains the desired physical information about the probed system.

\begin{figure*}\centering\includegraphics[width=0.95\textwidth]{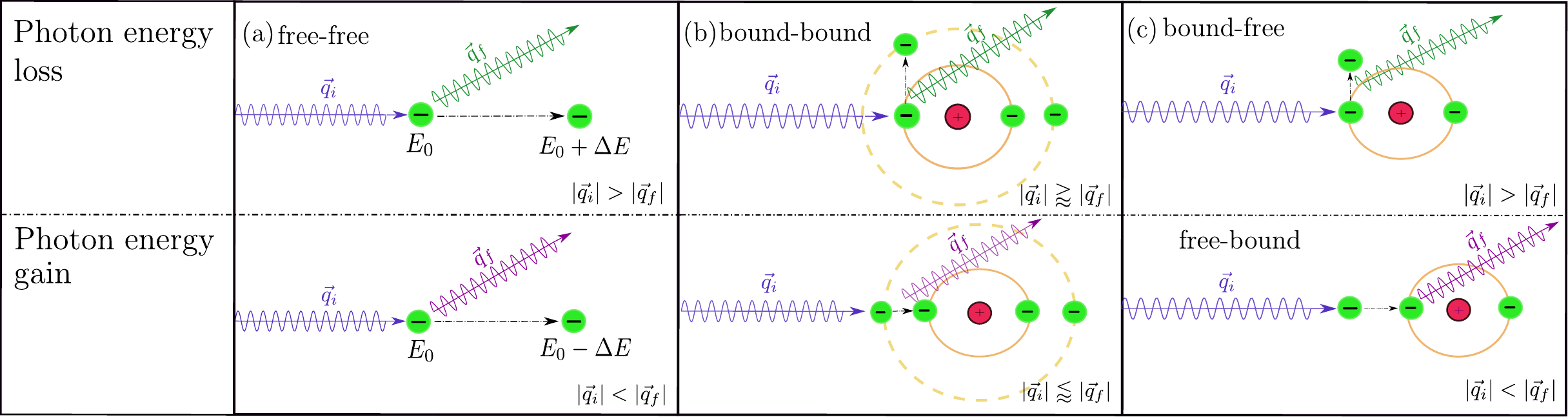}
\vspace*{0.2cm}
\caption{\label{fig:sketch} Illustration of the Chihara decomposition of the XRTS signal into different contributions involving effectively \emph{bound} and \emph{free} electrons. a) An incident photon (blue) is scattered on a free electron, leading to an energy loss (red-shifted) or energy gain (blue-shift) of the photon; b) scattering on a bound electron changing its state, also leading to either energy loss or gain; c) the top half depicts the scattering on a bound electron that gets lifted to the continuum due to the energy loss of the photon, i.e., a bound-free transition. The bottom half shows the hitherto unaccounted reverse process, i.e., the scattering on a free electron leading to an energy gain for the photon and transferring the electron to a bound state. Such \emph{free-bound} (FB) transitions are a particular feature of the complex physics emerging at WDM conditions.
}
\end{figure*}

Since the deconvolution of $I(\mathbf{q},\omega)$ to recover the dynamic structure factor is generally rendered unstable by noise in the experimental data, the standard approach for the interpretation of an XRTS measurement is to 1) construct a model for $S_{ee}(\mathbf{q},\omega)$ where unknown variables such as the temperature are being treated as fit parameters, 2) convolve the model with $R(\omega)$, and 3) determine the model parameters by finding the values such that the convolved model best fits the measured intensity signal.
On the one hand, this approach is, in principle, capable of giving access to a variety of properties, such as the equation of state (EOS)~\cite{Regan_PRL_2012,Falk_PRE_2013,Falk_PRL_2014}, and the thus inferred information constitutes constraints for EOS tables that impact a gamut of applications relevant to ICF~\cite{PhysRevLett.107.115004,hu_ICF,Hurricane_Nature_2014}, materials science, and astrophysical models~\cite{drake2018high}.
On the other hand, it is also clear that the inferred parameters can strongly depend on the employed model, which are typically based on a number of assumptions, such as the decomposition of the electronic orbitals into \emph{bound} and \emph{free} populations within the widely used Chihara ansatz~\cite{Chihara_1987,Gregori_PRE_2003} (see Fig.~\ref{fig:sketch} and its discussion below).
While the Chihara ansatz is exact in principle, its practical implementation in computationally efficient models constitutes a source of systematic error.
First-principles models for the dynamic structure factor overcome this~\cite{Baczewski_PRL_2016}, 
but the comparably large computational cost might render them impractical. 

In fact, the challenges of modelling XRTS signals directly reflect the notorious difficulty of finding a rigorous theoretical description of WDM. Indeed, a key feature of WDM is the intriguingly intricate interplay of a variety of physical effects, including quantum degeneracy (e.g.~Pauli blocking), Coulomb coupling, and strong thermal excitations~\cite{wdm_book,new_POP,Dornheim_review}. 
For temperatures comparable to the energies of bound states, the thermal excitation of electrons out of these states into free states enables x-ray scattering processes in which a free electron is de-excited back into an empty bound state, transferring energy and momentum to a scattered photon.

In the present work, we demonstrate that accounting for these scattering processes in interpreting XRTS experiments~\cite{Tilo_Nature_2023,kraus_xrts} improves fits based on the Chihara decomposition and brings inferred temperatures into better agreement with model-free temperature estimates~\cite{Dornheim_T_2022,Dornheim_T2_2022}.
The reinterpretation of these experiments has thus observed free-bound (FB) transitions in hard x-ray scattering for the first time.
As we explain below, such FB transitions are a distinct feature of WDM and vanish in the limits of cold and hot temperatures. 
Moreover, we show that the incorporation of FB transitions into the nearly universally used Chihara based interpretation of XRTS experiments has a direct and substantial impact on the inferred parameters, which is of immediate consequence for EOS measurements and the wide range of properties inferred from XRTS experiments.

\textbf{Idea.} The dynamic structure factor can be conveniently expressed in its exact spectral representation as~\cite{quantum_theory} 
\begin{eqnarray}\label{eq:spectral}
S_{ee}(\mathbf{q},\omega) = \sum_{m,l} P_m \left\|{n}_{ml}(\mathbf{q}) \right\|^2 \delta(\omega - \omega_{lm})\ ,
\end{eqnarray}
i.e., as a sum over all possible transitions between the eigenstates $l$ and $m$ of the full electronic Hamiltonian, with ${n}_{ml}(\mathbf{q})$ being the transition element induced by a density fluctuation of wave vector $\mathbf{q}$, $\omega_{lm}=(E_l-E_m)/\hbar$ the energy difference, and $P_m$ the occupation probability of the initial state $m$. Evidently, $S_{ee}(\mathbf{q},\omega)$ describes transitions where the scattered photon has lost energy to (gained energy from) the electronic system for $\omega>0$ ($\omega<0$). It is easy to see that the ratio of energy gain to energy loss is given by the simple detailed balance relation $S_{ee}(\mathbf{q},-\omega)/S_{ee}(\mathbf{q},\omega)=e^{-\hbar\omega/k_\textnormal{B}T}$ in thermodynamic equilibrium~\cite{quantum_theory,DOPPNER2009182}.

The basic idea behind the Chihara ansatz is the decomposition into scattering events involving effectively \emph{bound} and \emph{free} electrons. The total electronic dynamic structure factor is then given by $S_\textnormal{ee}(\mathbf{q},\omega) = S_\textnormal{FF}(\mathbf{q},\omega) + S_\textnormal{BB}(\mathbf{q},\omega) + S_\textnormal{BF}(\mathbf{q},\omega)$, and the individual components are explained in Fig.~\ref{fig:sketch}. Specifically, the first two terms describe transitions between two states where the scattered electron remains either free (free-free transitions) or transits between two bound states (bound-bound transitions).
We note that both types of transition can result either in an energy gain or an energy loss of the photon. Finally, $S_\textnormal{BF}(\mathbf{q},\omega)$ describes transitions of bound electrons into the continuum (bound-free transitions), cf.~Fig.~\ref{fig:sketch}c).
Yet Eq.~(\ref{eq:spectral}) directly implies that the reverse process is also possible: a FB transition, where the scattering of a photon on an initially free electron causes it to transition into a bound state, resulting in a contribution to $S_{ee}(\mathbf{q},\omega)$ for $\omega<0$. This contribution has been largely ignored in the XRTS-related WDM literature~\cite{Gregori_PRE_2003,kraus_xrts,siegfried_review}, even though it is directly analogous to the scattering processes that give rise to the blue-shifted plasmon peaks essential to XRTS thermometry.
We will demonstrate that its impact on the analysis of XRTS measurements can be substantial.

While there are many contexts in physics and chemistry in which electronic transitions between bound and free states contribute to some observable process (e.g., radiative cooling/recombination in plasmas~\cite{raymond1976radiative}, solids~\cite{van1954photon}, or ultracold gases~\cite{thorsheim1987laser}), in the context of XRTS these transitions are signatures of the WDM regime. At low temperatures ($k_\textnormal{B}T\ll E_\textnormal{F}$ with $E_\textnormal{F}$ being the Fermi energy), the probability to find an electron in any excited free state goes to zero. This leads to an exponential damping of the FB contribution by the detailed-balance factor, $S_\textnormal{FB}(\mathbf{q},\omega)=e^{-\hbar\omega/k_\textnormal{B}T}S_\textnormal{BF}(\mathbf{q},\omega)$.
Conversely, a given sample can be expected to be fully ionized in the hot-dense matter regime, such that bound-free transitions are not possible: i.e. $S_\textnormal{BF}(\mathbf{q},\omega)=S_\textnormal{FB}(\mathbf{q},\omega)=0$. In other words, the appearance of FB transitions in x-ray scattering is directly interwoven with partial ionization, which is a key feature of WDM systems~\cite{wdm_book}.



\begin{figure}\centering
\includegraphics[width=0.48\textwidth]{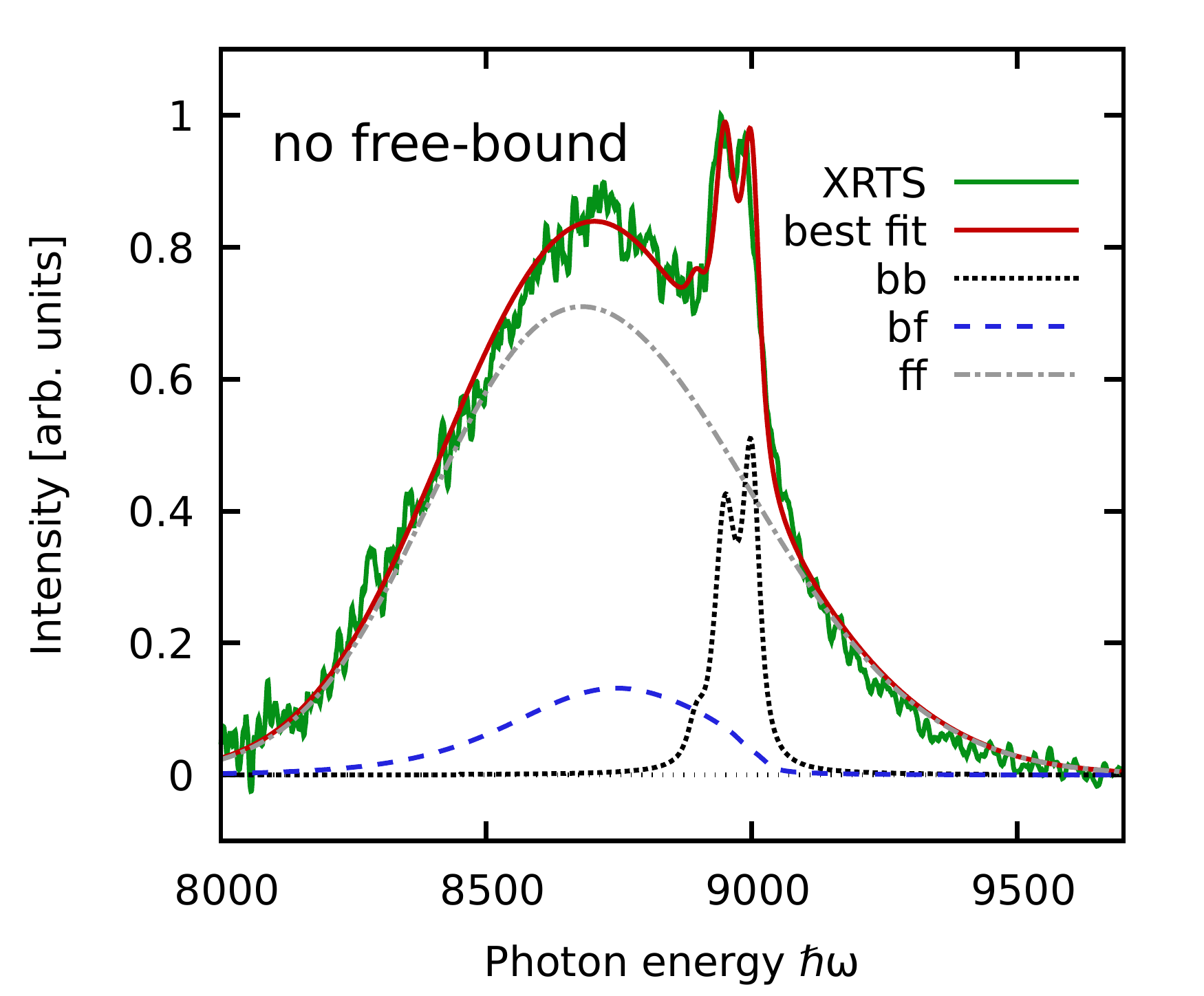}\\\vspace*{-0.5cm}
\includegraphics[width=0.48\textwidth]{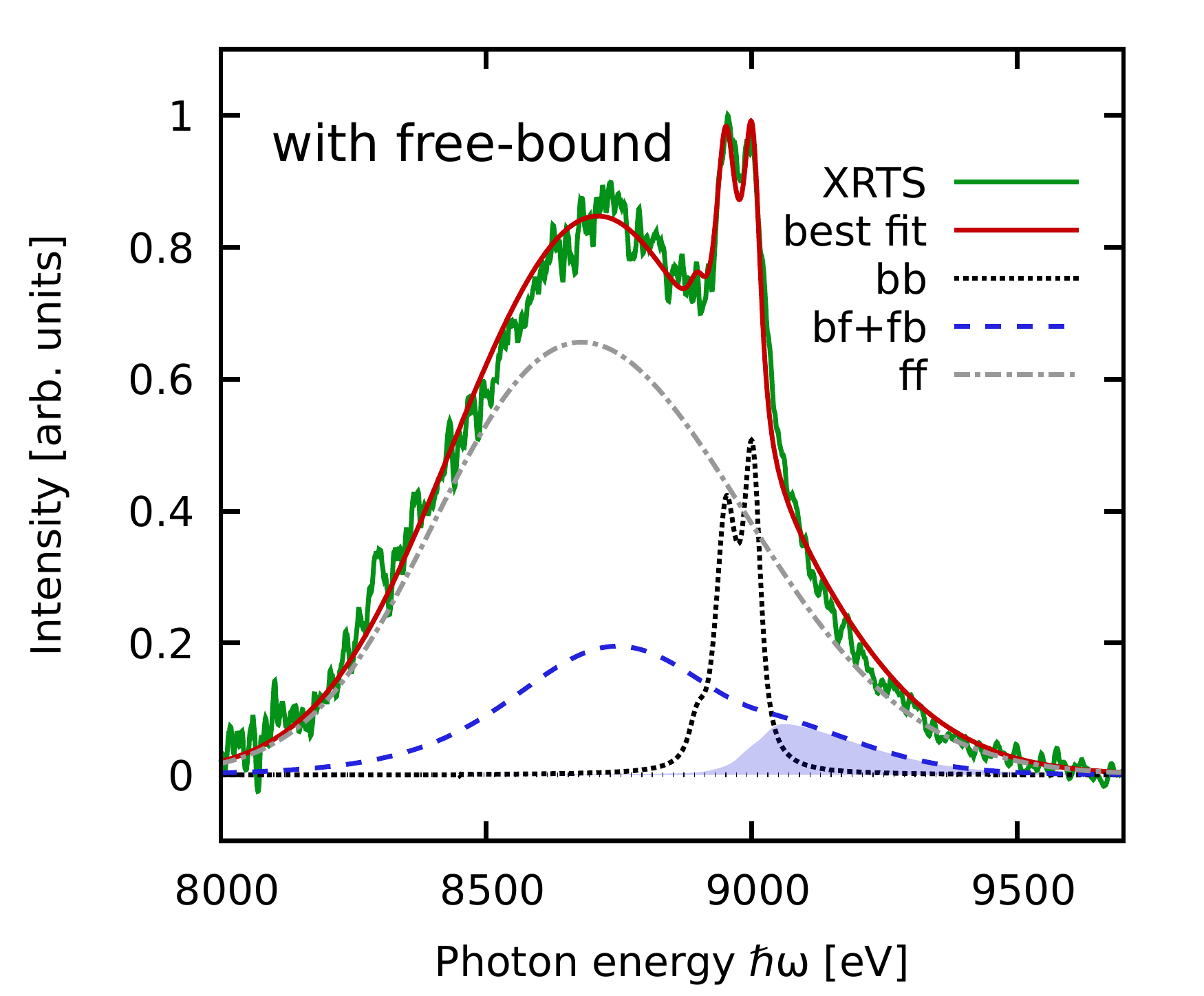}
\caption{\label{fig:NIF_Tilo} 
XRTS scattering intensity (solid green) of a Be ICF experiment at NIF ({N170214}) by D\"oppner \emph{et al.}~\cite{Tilo_Nature_2023} with a scattering angle of $\theta=120^\circ$ ($q=7.89\,$\AA$^{-1}$). The solid red line shows the best fit, and the dotted black, dashed blue, and dash-dotted grey lines show the corresponding components from the Chihara decomposition illustrated in Fig.~\ref{fig:sketch}. Top: Chihara fit without taking into account the physically mandated \emph{free-bound} transitions, with parameters taken from the original Ref.~\cite{Tilo_Nature_2023}. Bottom: Improved fit from the present work, with the FB contribution being indicated by the shaded blue area.
}
\end{figure}

\begin{figure}\centering
\includegraphics[width=0.48\textwidth]{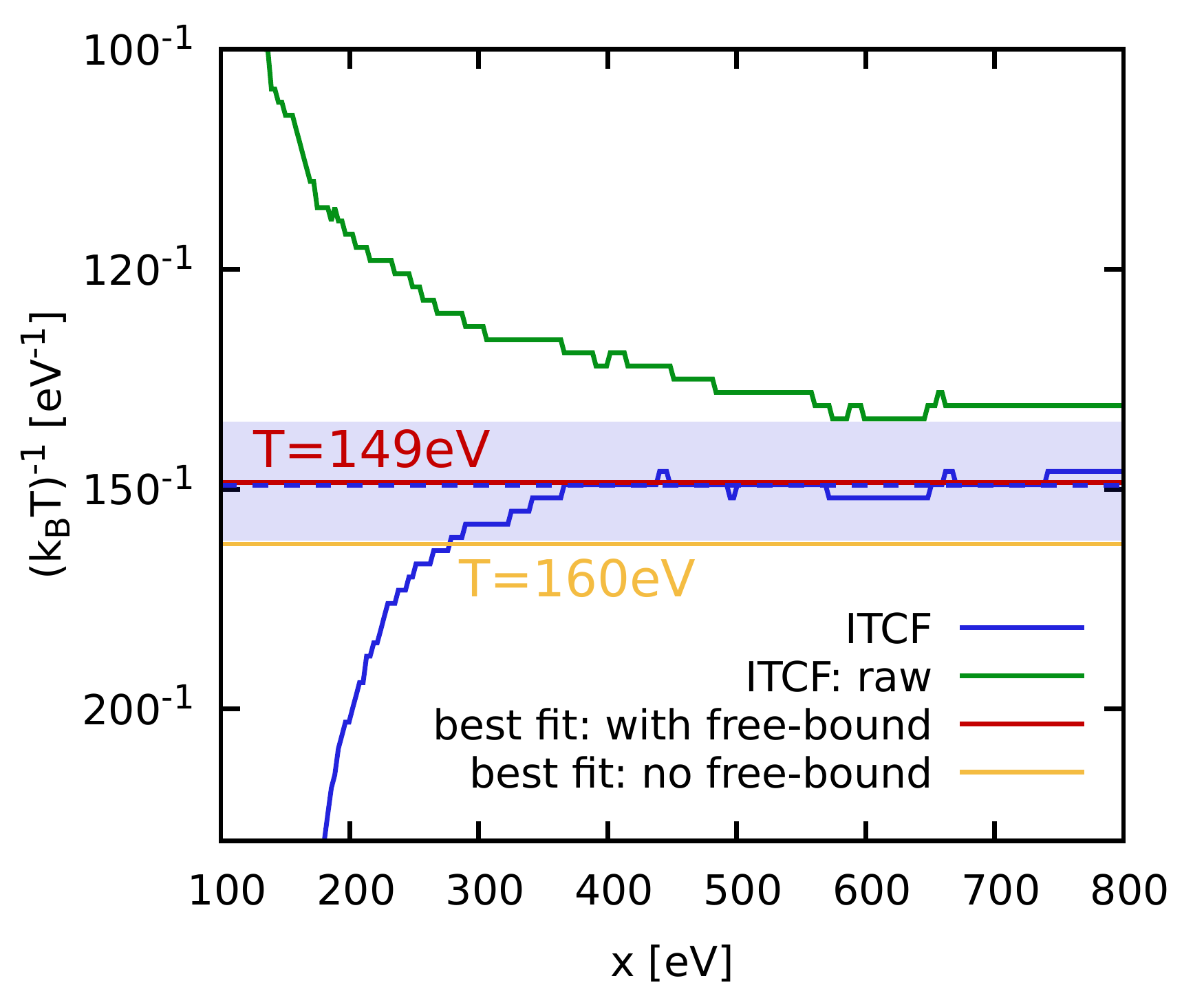}
\caption{\label{fig:NIF_Tilo_T} Convergence of the extracted temperature for the spectrum shown in Fig.~\ref{fig:NIF_Tilo} using the ITCF thermometry technique~\cite{Dornheim_T_2022,Dornheim_T2_2022} with the integration range $x$, see Eq.~(\ref{eq:truncated}) in the Methods section. Blue: ITCF estimate taking into account the instrument function $R(\omega)$; green: raw ITCF estimate neglecting $R(\omega)$; red: best fit properly taking into account free-bound transitions; orange: best fit from the original Ref.~\cite{Tilo_Nature_2023} without including the free-bound term.
}
\end{figure} 

\textbf{Results.} To demonstrate the practical importance and observation of FB transitions, we first consider a recent experiment at the NIF~\cite{Tilo_Nature_2023} where beryllium capsules were compressed using 184 of the 192 laser beams. An additional laser-driven x-ray backlighter source~\cite{MacDonald_POP_2022} (with zinc He-$\alpha$ lines at $8950\,$eV and $8999\,$eV) was used as a probe for the XRTS measurement, and the results are shown in Fig.~\ref{fig:NIF_Tilo}. The top panel shows the original analysis by D\"oppner et al.~\cite{Tilo_Nature_2023} where the solid green line depicts the measured XRTS signal, and the solid red line corresponds to the best fit. In addition, the dotted black, dash-dotted grey, and dashed blue lines show the respective components from bound-bound, free-free, and bound-free transitions. Here, the bound-free contribution violates detailed-balance and does not extend to the up-shifted part of the spectrum.

In the bottom panel, we redo this analysis and take into account the  FB contribution. 
We find that the free-bound component (shaded blue) to the total XRTS intensity has a substantial and significant weight, which means that the present analysis constitutes a rigorous observation of this effect in WDM. 
Furthermore, the inclusion of FB transitions fixes the previously violated, but physically mandated detailed balance condition. 
This is particularly relevant for the description of the up-shifted part of the spectrum.
In the original analysis~\cite{Tilo_Nature_2023}, the agreement between fit and measurement for $\omega>\omega_0$ was fully determined by the free-free part of the full spectrum, which is unphysical and necessarily leads to inconsistencies with the corresponding down-shifted side. The best fit shown in the top panel of Fig.~\ref{fig:NIF_Tilo} thus constitutes a compromise with a nominal temperature of $T=160\,$eV.
Incorporating the FB contribution removes this inconsistency, thereby lowering the extracted temperature to $T=149\,$eV when FB transitions are included.

While the rigorous benchmarking of such a Chihara-based analysis had been precluded by the lack of exact simulation tools in the past, this bottleneck has recently been partially lifted in Refs.~\cite{Dornheim_T_2022,Dornheim_T2_2022} where a highly accurate and model-free method to extract the temperature from XRTS measurements of WDM has been introduced. This methodology is based on Feynman's acclaimed imaginary-time path-integral formulation of statistical mechanics, and a brief introduction to the underlying idea of imaginary-time correlation function (ITCF) thermometry is presented in the \emph{Methods} section.
The results for the temperature analysis are shown in Fig.~\ref{fig:NIF_Tilo_T}, where the $y$-axis corresponds to the inverse temperature $\beta=1/k_\textnormal{B}T$ and the $x$-axis to the symmetrically truncated integration range, which is a convergence parameter for the ITCF approach. Moreover, the blue (green) line has been obtained  taking into account (not taking into account) the source and instrument function $R(\omega)$, and we find a clear onset of convergence for $x\gtrsim350\,$eV.

The temperature extracted from the best fit by D\"oppner \emph{et al.}~\cite{Tilo_Nature_2023} (horizontal orange line) is significantly larger than the ITCF estimate $T = 149\pm10\,$eV (blue line/shaded region). The physically consistent fit including FB transitions (horizontal red line), on the other hand, is in excellent agreement with the model-free ITCF result. A corresponding analysis of a second XRTS spectrum for Be that has been obtained during the same NIF shot, but at an earlier time (and, thus, an overall lower temperature) is presented in the Methods section.

\begin{figure}\centering
\includegraphics[width=0.48\textwidth]{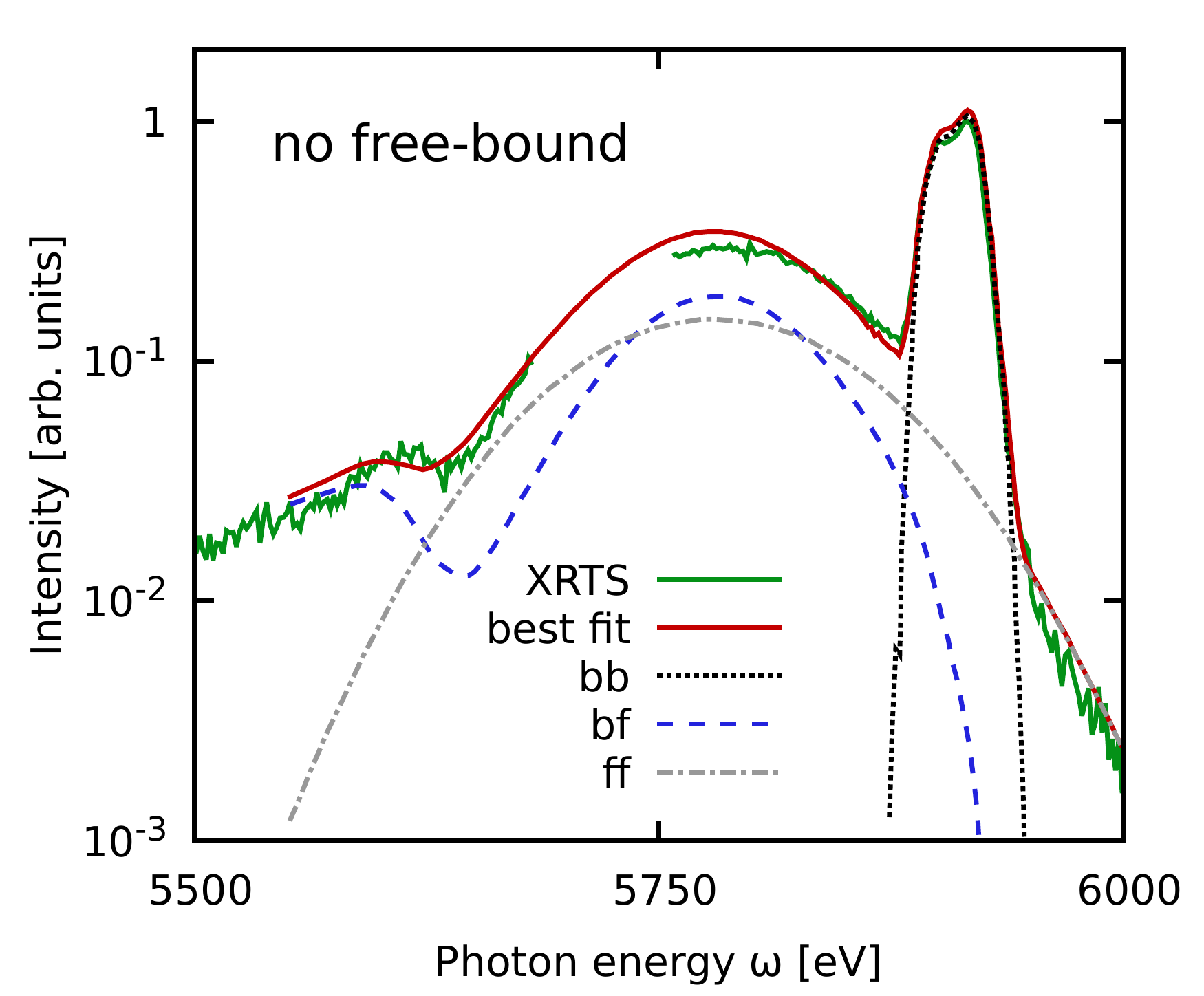}\\\vspace*{-0.5cm}
\includegraphics[width=0.48\textwidth]{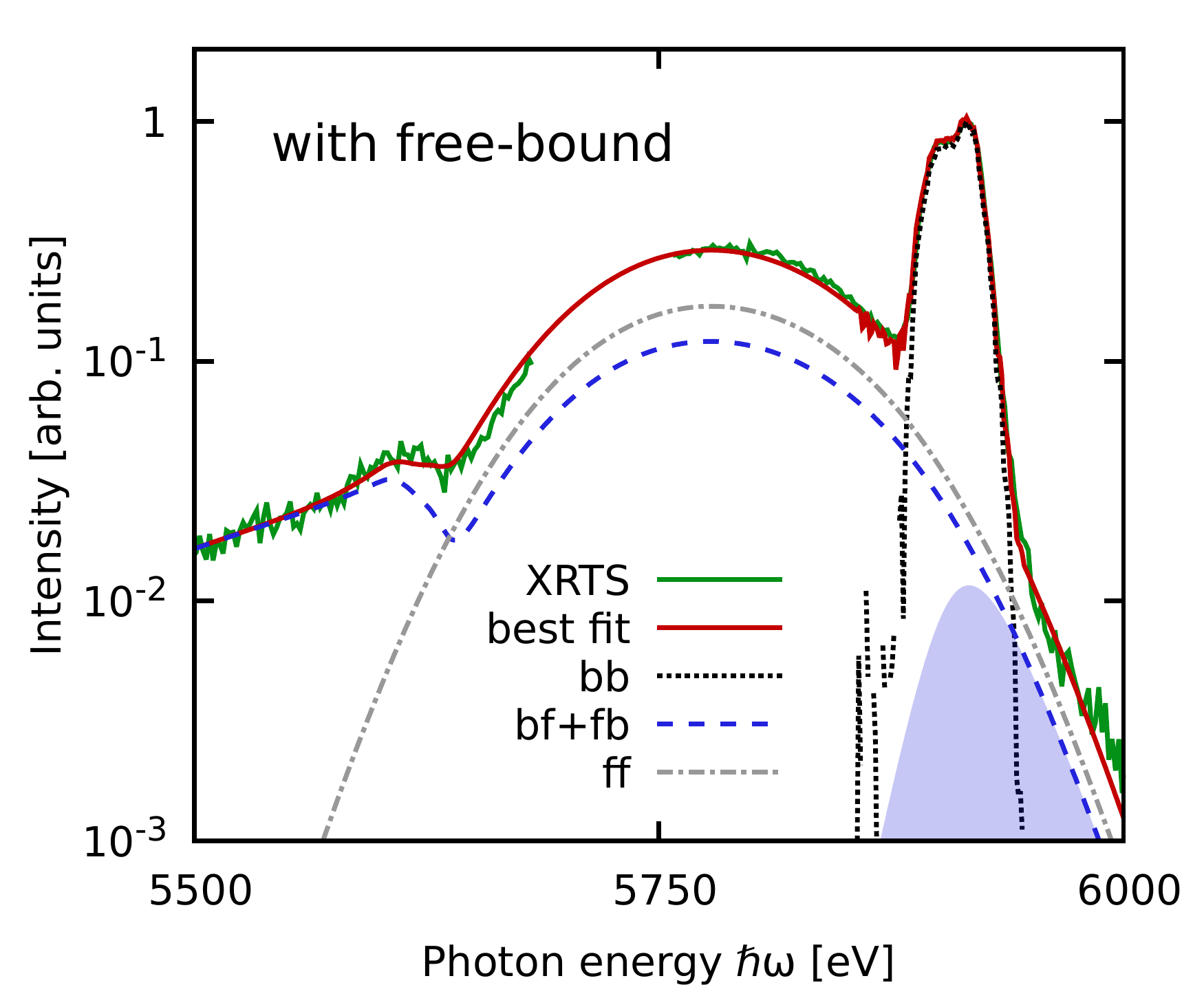}
\caption{\label{fig:carbon} 
XRTS scattering intensity (solid green) of isochorically heated graphite at a scattering angle of $\theta=160^\circ$ ($q=5.9\,$\AA$^{-1}$) obtained by Kraus \emph{et al.}~\cite{kraus_xrts} at LCLS. The solid red line shows the best fit, and the solid black, dashed blue, and dash-dotted grey lines show the corresponding components from the Chihara decomposition illustrated in Fig.~\ref{fig:sketch}. Top: Chihara fit without taking into account the physically mandated \emph{free-bound} transitions, with parameters taken from the original Ref.~\cite{kraus_xrts}. Bottom: Improved fit from the present work, with the FB contribution being indicated by the shaded blue area.
}
\end{figure}

As a second example, we consider an XRTS measurement of isochorically heated graphite that was performed by Kraus \emph{et al.}~\cite{kraus_xrts} at LCLS. The corresponding XRTS signal is shown as the solid green curve in Fig.~\ref{fig:carbon}, and the top panel corresponds to the original analysis from Ref.~\cite{kraus_xrts}. Evidently, the bound-free feature due to the excitation of weakly bound L-shell electrons constitutes the dominant contribution to the inelastic feature around $\omega=5800\,$eV; it is, however, absent from the up-shifted part. As a consequence, the thus constructed model for the full XRTS intensity strongly violates the detailed balance between energy loss and energy gain. Moreover, the up-shifted feature is fitted entirely by the free-free contribution, leading to similar inconsistencies as reported for the NIF Be shot investigated in Fig.~\ref{fig:NIF_Tilo} above.
In the bottom panel of Fig.~\ref{fig:carbon}, we show our present analysis based on the improved, physically consistent Chihara model that includes FB transitions; the latter are highlighted as the shaded blue area in the up-shifted part of the spectrum. Our new fit confers a substantial improvement compared to the original fit over the entire frequency range. The temperature that we have extracted from the best fit is given by $T=16.6\,$eV, which is in excellent agreement with the ITCF-based value of $T=18\pm2\,$eV given in Ref.~\cite{Dornheim_T2_2022} that has been obtained without any model assumptions; see the \emph{Methods} section for additional details. In contrast, the best fit by Kraus \emph{et al.}~\cite{kraus_xrts} gives $T=21.7\,$eV, and thus significantly overestimates the temperature. This again highlights the importance of FB transitions for the determination of the equation-of-state of materials at WDM conditions based on XRTS measurements.

\begin{figure}\centering\vspace*{-0.35cm}
\includegraphics[width=0.45\textwidth]{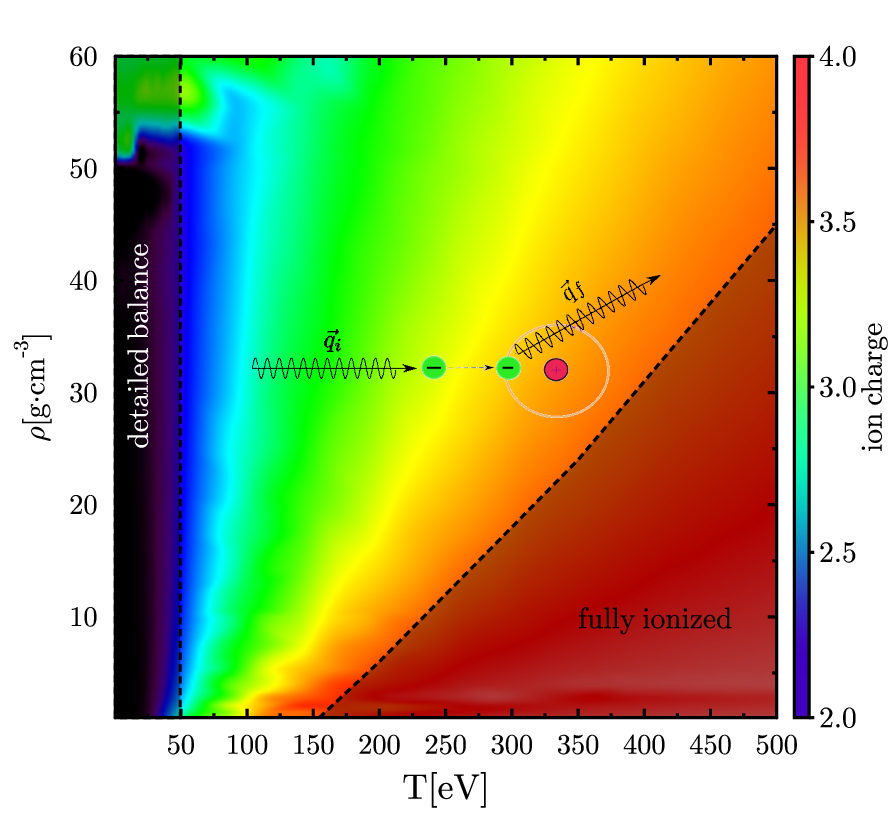}\\\vspace*{-0.55cm}
\includegraphics[width=0.44\textwidth]{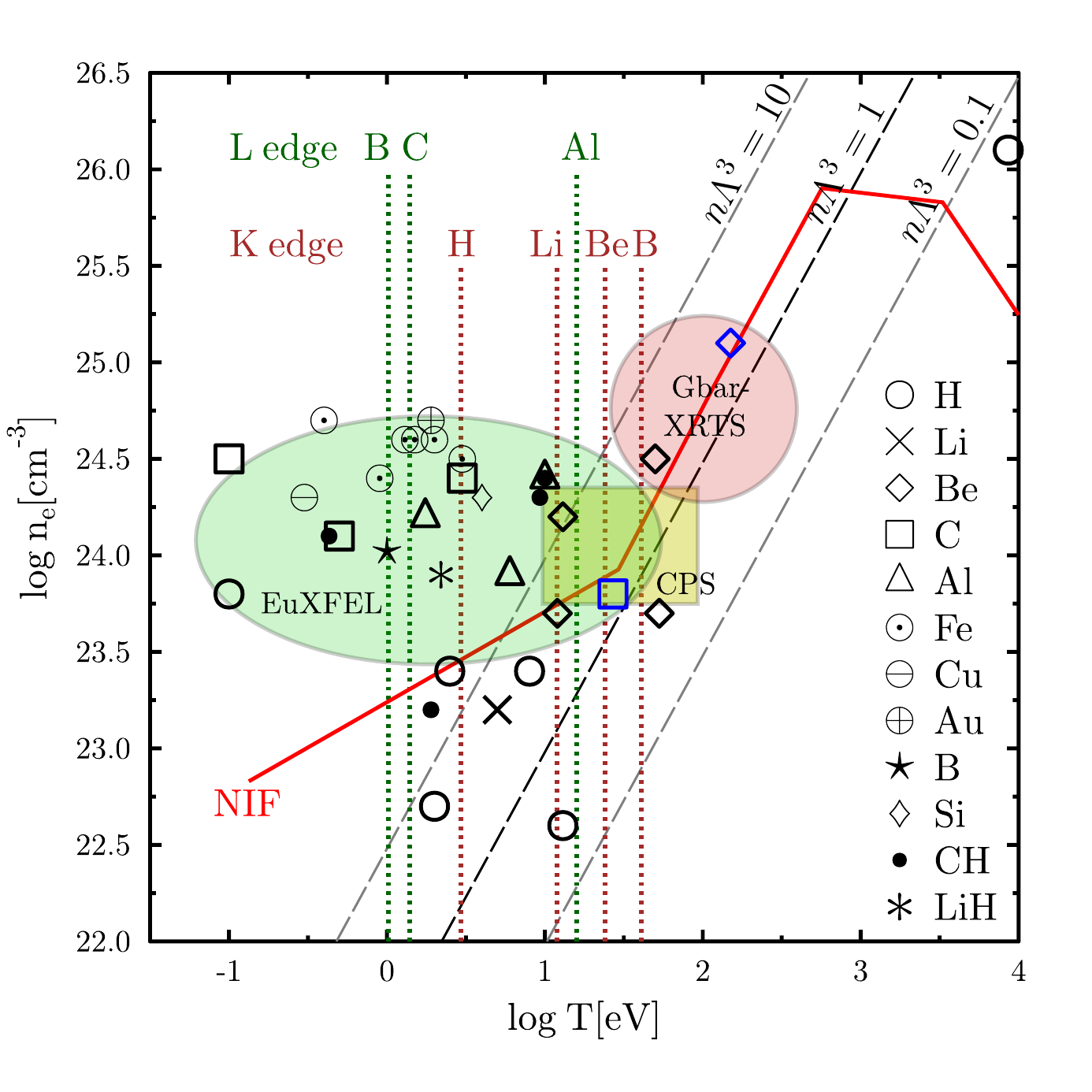}\vspace*{-0.35cm}
\caption{\label{fig:jan} 
Top: Ionization degree of beryllium computed by FLYCHK~\cite{CHUNG20053}. 
FB transitions are important for partial ionization, i.e., in the bright area around the center; they are suppressed by the detailed balance relation for $T\lesssim50\,$eV (left), and by full ionization for $T\gtrsim150-400\,$eV depending on $\rho$ (bottom right). Bottom: overview of selected XRTS experiments with different elements (see point style) in the WDM regime, shown in the temperature-number density plane. The vertical dotted green (red) lines indicate the minimum temperature above which FB transitions become significant for the respective L edge (K edge). The shaded green, yellow, and red areas correspond to experimental capabilities at the European XFEL~\cite{Tschentscher_2017}, the brandnew Colliding Planar Shock platform (CPS) at NIF~\cite{MacDonald_POP_2023}, and the Gigabar-XRTS platform at NIF~\cite{Tilo_Nature_2023}. The red line shows a simulated NIF ICF implosion path.
}
\end{figure}

\textbf{Discussion.} In this work, we have unambiguously demonstrated the importance of including \emph{free-bound} transitions in models of XRTS signals, both to restore physically mandated detailed balance and reduce sources of systematic error. 
The prominence of signatures of these processes is a direct consequence of the complex interplay of various physical effects emerging in WDM; it vanishes both in the limits of low temperatures and in the hot dense matter regime.
This is illustrated in the top panel of Fig.~\ref{fig:jan}, where we show the ionization degree of Be as a function of temperature and mass density. In particular, FB transitions are significant in the bright area around the center, i.e., above a threshold of $T=50\,$eV due to the detailed balance factor, and below full ionization at around $T=150-400\,$eV (bottom right), depending on $\rho$.

To demonstrate the importance of this effect, we have re-analyzed two representative XRTS experiments with WDM: the recent investigation of strongly compressed Be at the NIF by D\"oppner \emph{et al.}~\cite{Tilo_Nature_2023}, and an investigation of isochorically heated graphite at LCLS by Kraus~\emph{et al.}~\cite{kraus_xrts}. In both cases, we have found that including free-bound transitions leads to a reduction of the inferred temperature, and an improved agreement between theory and measurement. These findings have been further substantiated by the excellent agreement of the thus extracted temperature with the independent and model-free ITCF thermometry technique developed in Refs.~\cite{Dornheim_T_2022,Dornheim_T2_2022}.
To put our results into the proper context, we show an overview of a number of XRTS experiments with different elements in the WDM regime in Fig.~\ref{fig:jan}b). The different symbols distinguish different materials, and the dotted vertical green (red) lines show the temperature above which FB transitions become important for the respective L edge (K edge). The experiments with Be (cf.~Fig.~\ref{fig:NIF_Tilo}) and carbon (cf.~Fig.~\ref{fig:carbon}), as reanalyzed here, are indicated by the blue diamond and square, respectively.
Furthermore, the shaded areas indicate the experimental capabilities at two key facilities. The shaded green bubble corresponds to the European XFEL~\cite{Tschentscher_2017}, which combines excellent beam properties with an unprecedented repetition rate for pump-probe experiments. The latter allows for highly accurate XRTS measurements over a dynamic range of at least three orders of magnitude in the intensity~\cite{Voigt_POP_2021}, which makes it possible to resolve FB transitions even for $T\sim10\,$eV.
The red circle outlines parameters realized at the NIF Gbar-XRTS platform~\cite{Tilo_Nature_2023}, where FB transitions indeed constitute a predominant feature; cf.~our re-analysis of the Be experiment by D\"oppner \emph{et al.}~\cite{Tilo_Nature_2023} in Fig.~\ref{fig:NIF_Tilo} above.
Of particular importance is the olive square corresponding to the new Colliding Planar Shock (CPS) platform at NIF~\cite{MacDonald_POP_2023}. It has been designed to realize very uniform conditions that are ideally suited for precision EOS measurements. Yet, the accurate determination of the latter requires FB transitions to be taken into account for the interpretation of XRTS experiments, as it can be clearly seen from Fig.~\ref{fig:jan}.

The work presented here will have a direct and profound impact on a number of research fields related to the study of matter under extreme conditions. XRTS constitutes a widely used method of diagnostics for WDM and beyond. This makes it particularly important for benchmarking EOS models~\cite{Falk_PRL_2014}, which constitute a key input for understanding astrophysical phenomena~\cite{drake2018high}, fusion applications~\cite{PhysRevLett.107.115004,hu_ICF}, and a plethora of other calculations~\cite{Militzer_PRE_2021}. The particular relevance of FB transitions to fusion applications is further substantiated by a simulated ICF implosion path that has been included as the solid red line in the bottom panel of Fig.~\ref{fig:jan}.

Furthermore, incorporating free-bound transitions into the Chihara model restores the exact detailed balance relation. This opens up the way for the systematic improvement of the underlying theoretical description of the individual components (cf.~Fig.~\ref{fig:sketch}), which, otherwise, might have been biased by the artificial distortion between the positive and negative frequency range; this would likely have prevented agreement between theory and experiment even for an exact description of $S_\textnormal{FF}(\mathbf{q},\omega)$, $S_\textnormal{BB}(\mathbf{q},\omega)$, and $S_\textnormal{BF}(\mathbf{q},\omega)$ in many cases.

A particularly enticing route for future research is given by the application of the improved, physically consistent Chihara model to emerging exact path integral Monte Carlo simulation results for warm dense hydrogen~\cite{Bohme_PRL_2022} and other light elements. These efforts will allow us to rigorously benchmark existing models for the individual contributions to $S_{ee}(\mathbf{q},\omega)$, and to guide the development of improved theories. Moreover, such a comparison will allow us to assess the conceptual validity of the decomposition into \emph{bound} and \emph{free} electrons on which the Chihara model is based, which will be of fundamental importance to our understanding of electron--ion systems in different contexts.




\backmatter
\section*{Methods}

\bmhead{Model-free imaginary-time correlation function temperature diagnostics}\label{sec:ITCF}

In the main text, we have demonstrated the improved accuracy of the Chihara fit including the \emph{free-bound} transitions by comparing the extracted temperature with the highly accurate and model-free imaginary-time correlation function (ITCF) thermometry technique~\cite{Dornheim_T_2022,Dornheim_T2_2022}.
The basic idea is to consider the two-sided Laplace transform of $S(\mathbf{q},\omega)$, which gives the imaginary-time version of the usual intermediate scattering function $F_{ee}(\mathbf{q},t)$~\cite{siegfried_review},
\begin{eqnarray}\label{eq:Laplace}
    F_{ee}(\mathbf{q},\tau) = \int_{-\infty}^\infty \textnormal{d}\omega\ S_{ee}(\mathbf{q},\omega) e^{-\tau\hbar\omega}\ .
\end{eqnarray}
Specifically, the time argument has been replaced by $t=-i\hbar\tau$, with $\tau\in[0,\beta]$ and $\beta=1/k_\textnormal{B}T$. From a mathematical perspective, both the $\tau$- and the $\omega$-representation are formally equivalent. It has recently been pointed out that working in the imaginary-time domain has a number of key advantages~\cite{Dornheim_T_2022,Dornheim_T2_2022,Dornheim_review}. First, the deconvolution with respect to the source-and-instrument function $R(\omega)$ becomes straightforward even in the presence of substantial noise in the experimental data. Second, the detailed balance relation connecting $S_{ee}(\mathbf{q},\omega)$ with $S_{ee}(\mathbf{q},-\omega)$ leads to the symmetry relation $F_{ee}(\mathbf{q},\tau)=F_{ee}(\mathbf{q},\beta-\tau)$, which always manifests as a minimum of $F_{ee}(\mathbf{q},\tau)$ around $\tau=\beta/2 =1/2T$. In other words, locating the minimum in the ITCF Eq.~(\ref{eq:Laplace}) gives one direct access to the temperature of an arbitrarily complex system without any simulations or approximations.

An additional difficulty is given by the necessarily finite spectral range of any experimental data set, whereas the evaluation of Eq.~(\ref{eq:Laplace}), in principle, would require an integration from negative to positive infinity. In practice, we compute the symmetrically truncated ITCF
\begin{eqnarray}\label{eq:truncated}
    F_x(\mathbf{q},\tau) = \int_{-x}^x\textnormal{d}\omega\ S_{ee}(\mathbf{q},\omega) e^{-\tau\hbar\omega}\ ,
\end{eqnarray}
and the convergence of the thus extracted temperature with the integration range $x$ is demonstrated in Fig.~\ref{fig:NIF_Tilo_T} in the main text.

\bmhead{Analysis of complementary NIF spectrum at lower temperature}\label{sec:low}



\begin{figure}\centering
\includegraphics[width=0.48\textwidth]{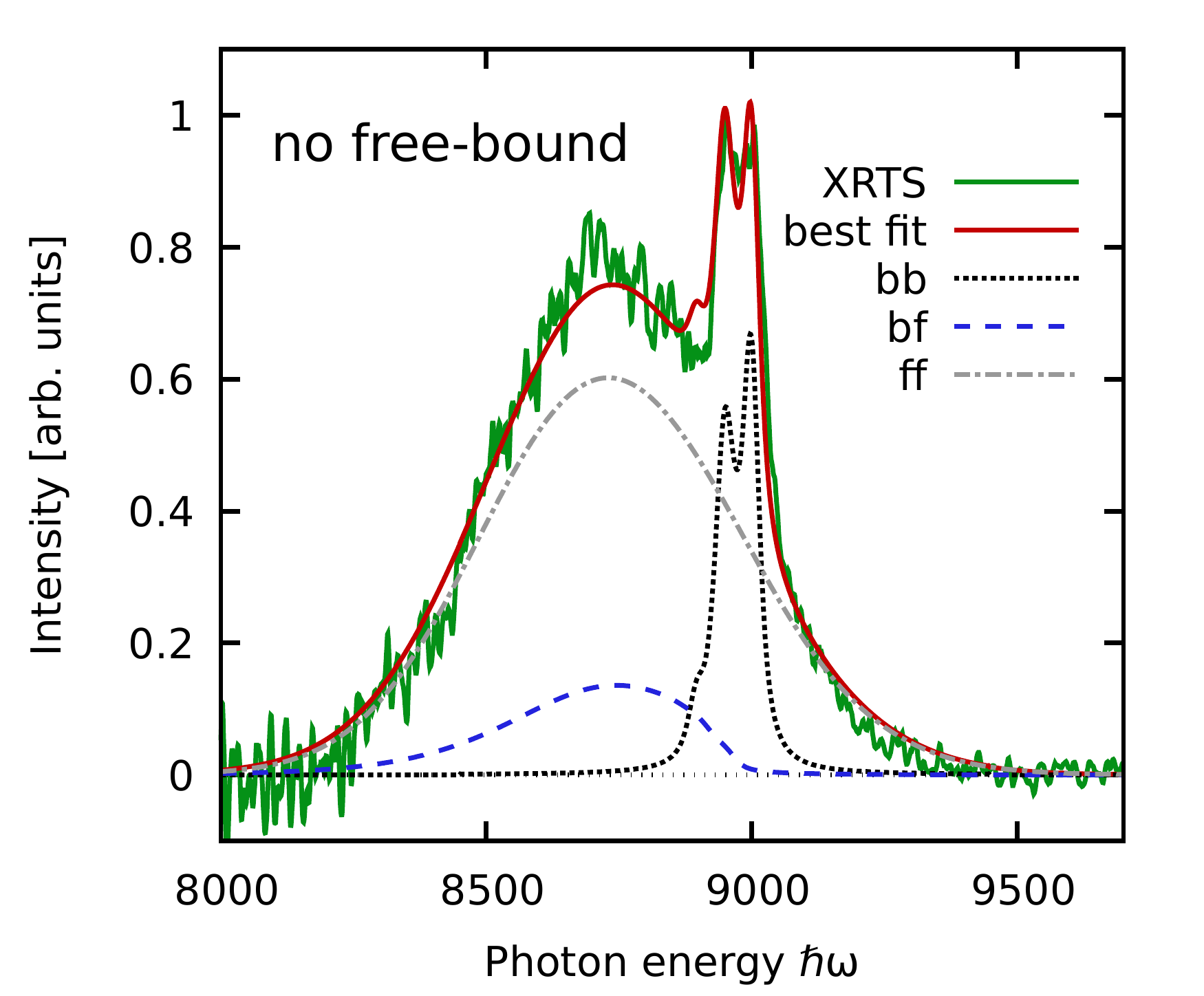}\\\vspace*{-0.5cm}
\includegraphics[width=0.48\textwidth]{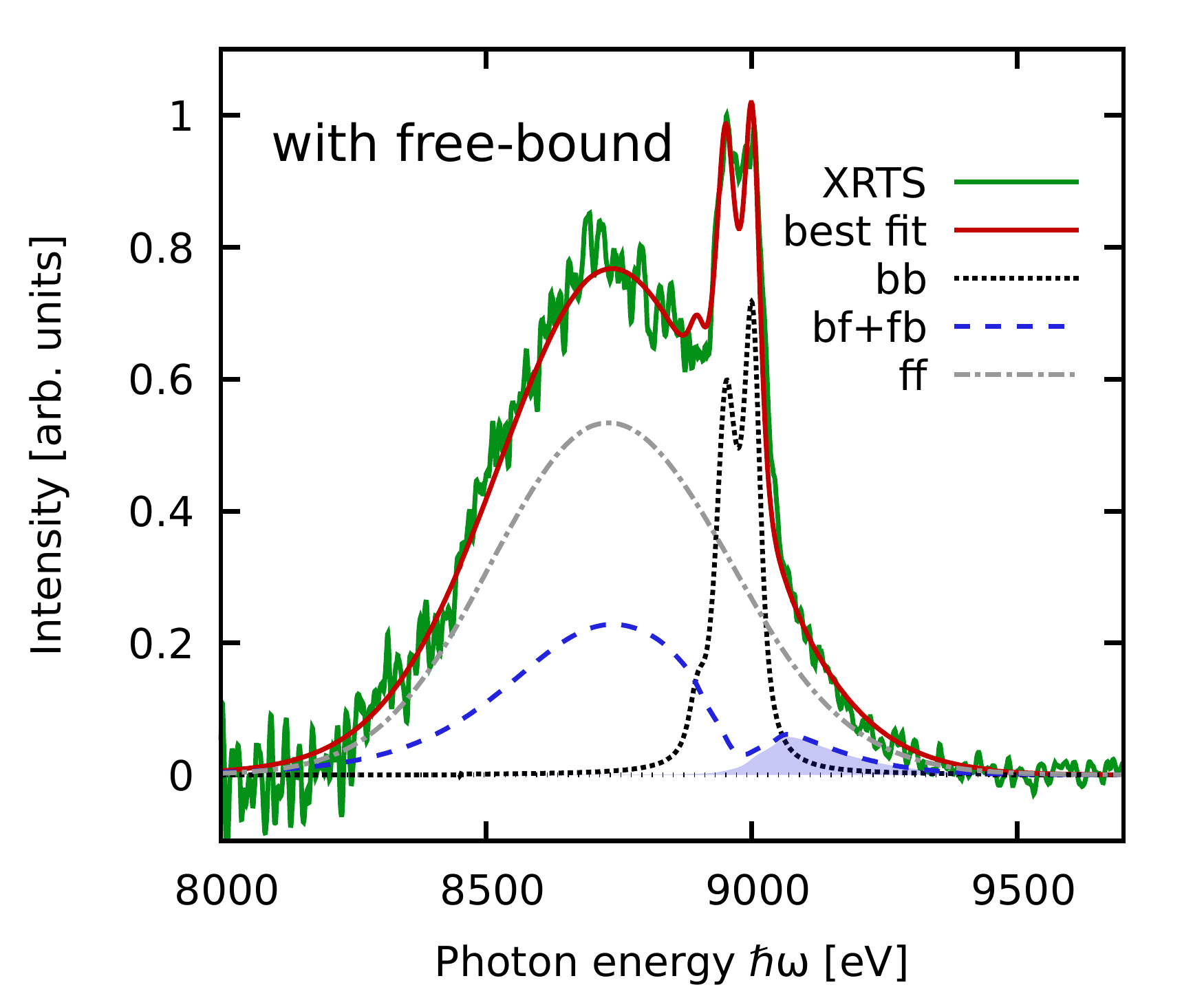}
\caption{\label{fig:NIF_Tilo_2} 
XRTS scattering intensity (solid green) of the same Be ICF experiment at NIF ({N170214}) by D\"oppner \emph{et al.}~\cite{Tilo_Nature_2023} shown in Fig.~\ref{fig:NIF_Tilo} in the main text, but at an earlier probe time where the capsule is less compressed and less heated. The solid red line shows the best fit, and the dotted black, dashed blue, and dash-dotted grey lines show the corresponding components from the Chihara decomposition illustrated in Fig.~\ref{fig:sketch}. Top: Chihara fit without taking into account the physically mandated \emph{free-bound} transitions, with parameters taken from the original Ref.~\cite{Tilo_Nature_2023}. Bottom: Improved fit from the present work, with the FB contribution being indicated by the shaded blue area.
}
\end{figure}

\begin{figure}\centering
\includegraphics[width=0.48\textwidth]{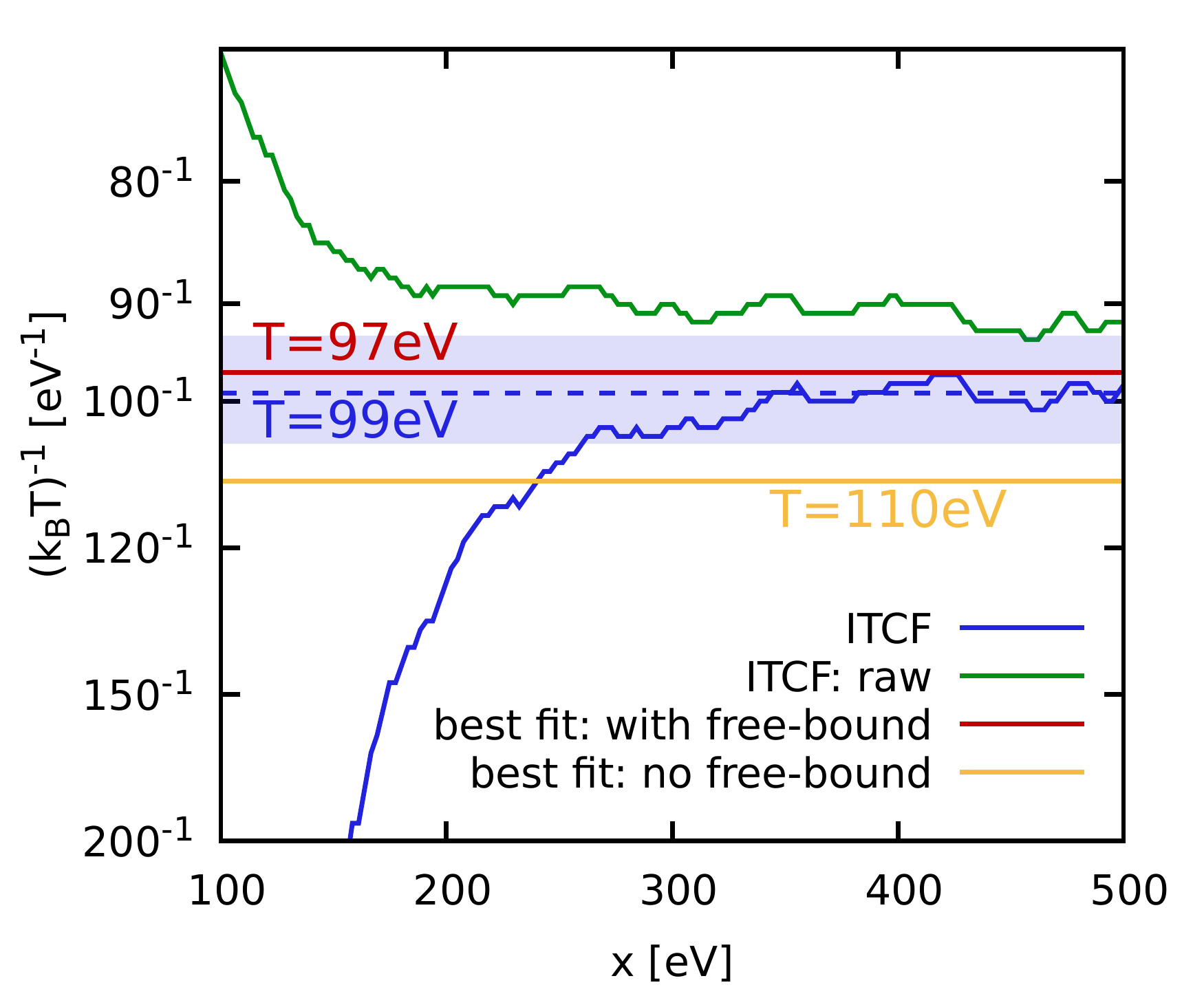}
\caption{\label{fig:NIF_Tilo_T_2} Convergence of the extracted temperature for the spectrum shown in Fig.~\ref{fig:NIF_Tilo_2} using the ITCF thermometry technique~\cite{Dornheim_T_2022,Dornheim_T2_2022} with the integration range $x$, see Eq.~(\ref{eq:truncated}) in the Methods section. Blue: ITCF estimate taking into account the instrument function $R(\omega)$; green: raw ITCF estimate neglecting $R(\omega)$; red: best fit properly taking into account free-bound transitions; orange: best fit from the original Ref.~\cite{Tilo_Nature_2023} without including the free-bound term.
}
\end{figure}

To further demonstrate the general nature of FB transitions for the interpretation of XRTS experiments probing WDM states, we re-analyze a second measurement of compressed Be in Fig.~\ref{fig:NIF_Tilo_2}. It has been obtained during the same experiment (N170214) as the spectrum shown in Fig.~\ref{fig:NIF_Tilo} in the main text, but at an earlier time, leading to a comparably reduced degree of compression and a lower temperature.
The top and bottom panels again show the Chihara-based analysis without and with including the FB contribution, and we find the same trends as in the previous examples. 

The corresponding ITCF analysis of the extracted temperature is shown in Fig.~\ref{fig:NIF_Tilo_T_2}. The best fit from the original Ref.~\cite{Tilo_Nature_2023} gives a temperature of $T=110\,$eV (orange), whereas our improved model that includes FB transitions lowers that value to $T=97\,$eV (red). The latter is in significantly better agreement with the properly deconvolved ITCF result of $T=99\pm7\,$eV (blue), as it is expected.

\bmhead{Impact of the instrument function model}\label{sec:SIF}

\begin{figure}\centering
\includegraphics[width=0.48\textwidth]{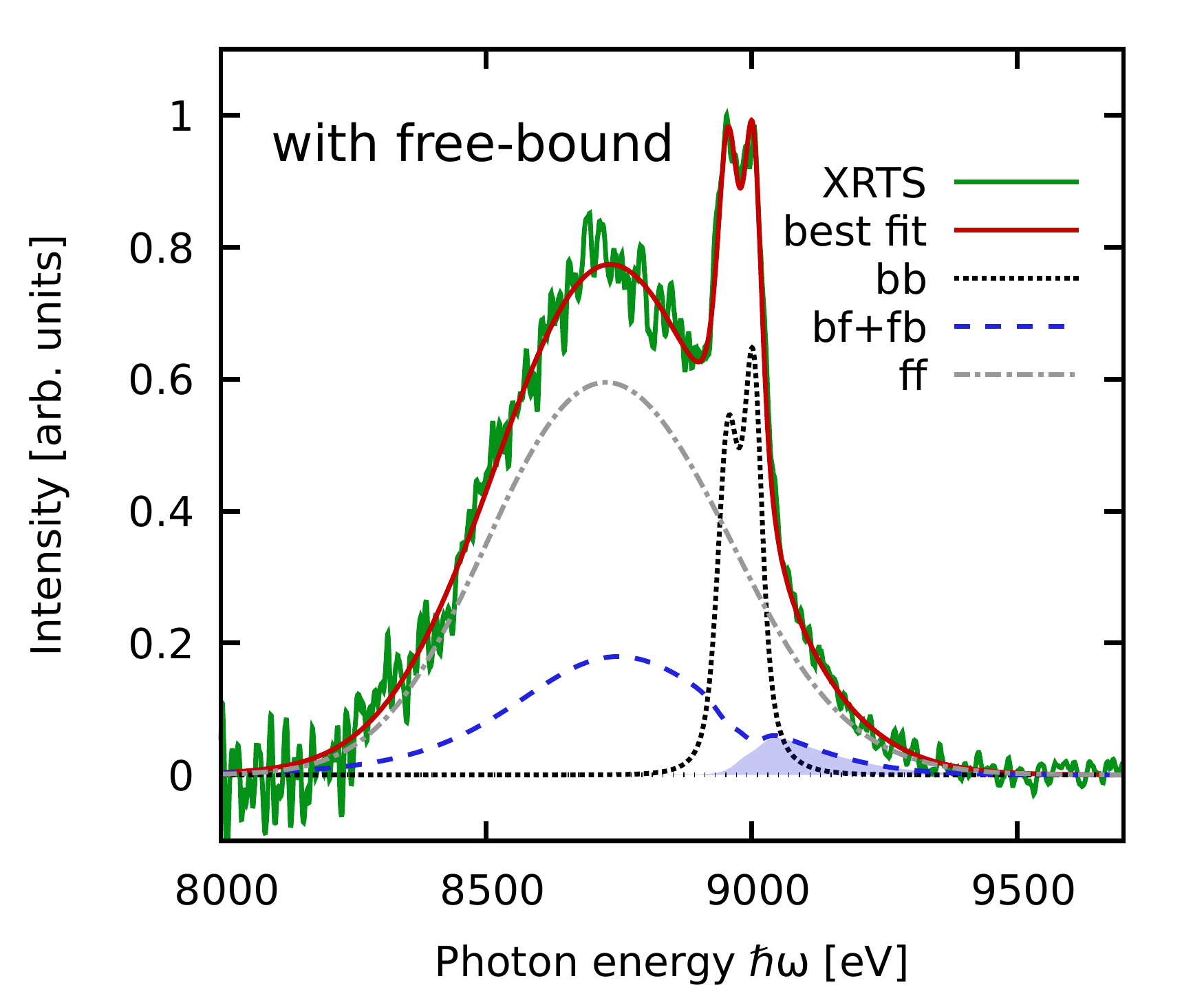}
\caption{\label{fig:instrument} 
Re-analyzing the XRTS experiment shown in Fig.~\ref{fig:NIF_Tilo_2} with a modified, empirical model for the combined source and instrument function.
}
\end{figure}

An important aspect of the rigorous interpretation of XRTS experiments is the characterization of the combined source and instrument function~\cite{MacDonald_POP_2022}. This holds both for the model-free ITCF method~\cite{Dornheim_T_2022,Dornheim_T2_2022} discussed above, and for the Chihara-based forward modelling approach that constitutes the focus of the present work.
In Fig.~\ref{fig:instrument}, we have repeated the analysis of the colder Be NIF spectrum shown in the bottom panel of Fig.~\ref{fig:NIF_Tilo_2}, but using an empirical model for the instrument function that has been obtained by fitting FLYCHK emission lines to the elastic feature directly extracted from the measured XRTS signal. The main differences to the source function model used in the original Ref.~\cite{Tilo_Nature_2023} are the absence of the left shoulder, and a somewhat less pronounced double peak structure.

\begin{figure}\centering
\includegraphics[width=0.48\textwidth]{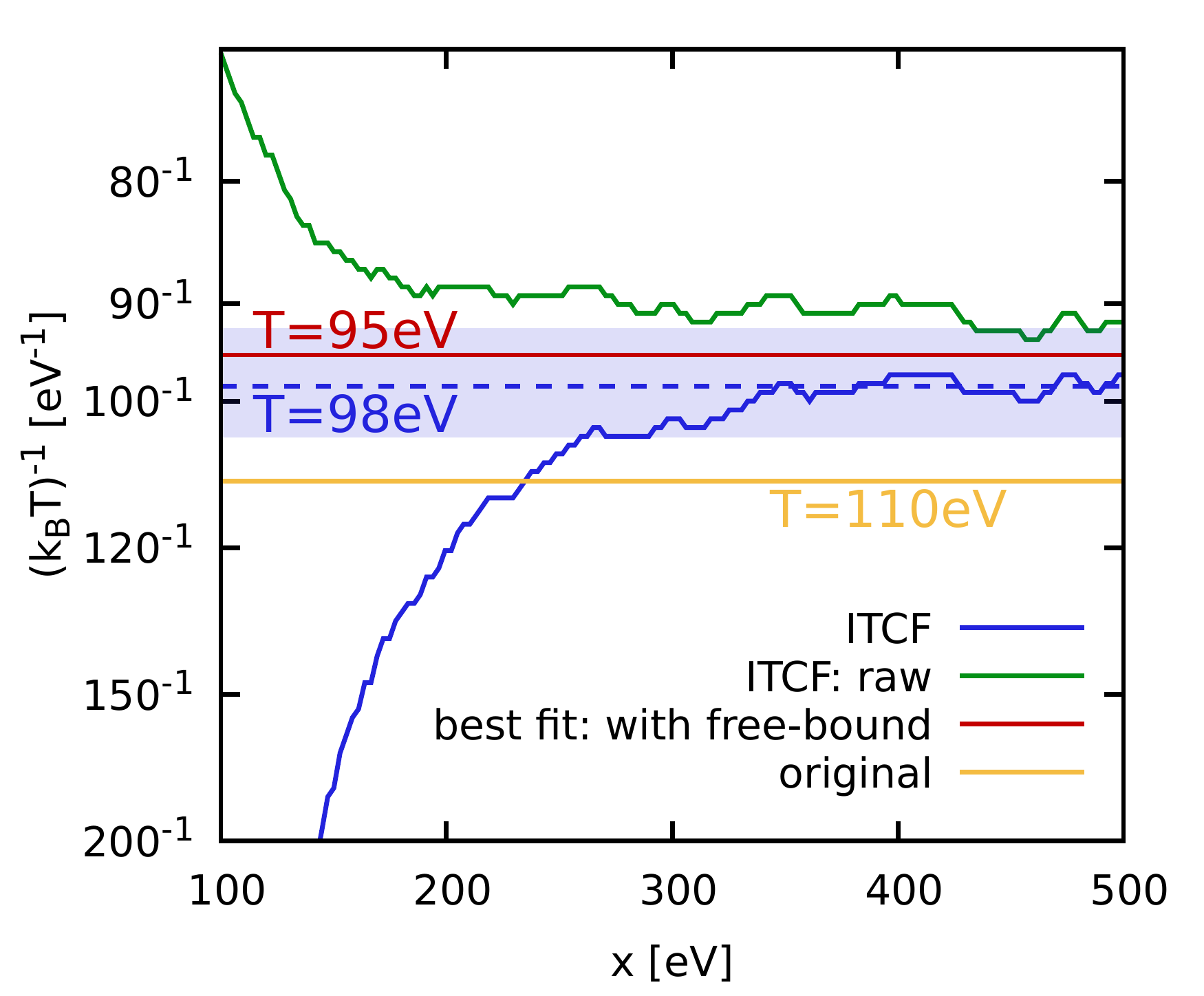}
\caption{\label{fig:instrument_tau} Convergence of the extracted temperature for the spectrum shown in Fig.~\ref{fig:instrument} using the ITCF thermometry technique~\cite{Dornheim_T_2022,Dornheim_T2_2022} with the integration range $x$, see Eq.~(\ref{eq:truncated}) in the Methods section. Blue: ITCF estimate taking into account the instrument function $R(\omega)$; green: raw ITCF estimate neglecting $R(\omega)$; red: best fit properly taking into account free-bound transitions; orange: best fit from the original Ref.~\cite{Tilo_Nature_2023} without including the free-bound term.
}
\end{figure}

Evidently, this empirical, data-driven source function leads to a substantially improved agreement between the Chihara model and the experimental data. The corresponding temperature analysis is shown in Fig.~\ref{fig:instrument_tau}, and we find the same trends as in the previous section.



This leads us to the following conclusions: i) residual deviations between the improved Chihara model that includes FB transitions and experimental measurements can be a consequence either of systematic inaccuracies in the individual components of the former, or of the employed model for the source and instrument function; ii) the quantification of uncertainties in the source function, and the systematic study of their impact on both forward modelling and ITCF based analysis frameworks constitutes an important task for future work; iii) the observation of the physically mandated FB transitions reported in the present work is very robust and does not depend on any particular model.

\section*{Data Availability}

The spectra generated from our improved Chihara model that includes free-bound transitions will be made openly available in the Rossendorf data repository (RODARE).




\bibliography{sn-bibliography}

\begin{thebibliography}{10}
\providecommand{\doi}[1]{\url{https://doi.org/#1}}
\bibcommenthead

\bibitem{wdm_book}
Graziani F, Desjarlais MP, Redmer R, Trickey SB, editors.
\newblock Frontiers and Challenges in Warm Dense Matter.
\newblock International Publishing: Springer; 2014.

\bibitem{drake2018high}
Drake RP.
\newblock High-Energy-Density Physics: Foundation of Inertial Fusion and
  Experimental Astrophysics.
\newblock Graduate Texts in Physics. Springer International Publishing; 2018.

\bibitem{Hatfield_Nature_2021}
Hatfield PW, Gaffney JA, Anderson GJ, Ali S, Antonelli L,
  Ba{\c{s}}e{\u{g}}mez~du Pree S, et~al.
\newblock The data-driven future of high-energy-density physics.
\newblock Nature. 2021 May;593(7859):351--361.
\newblock \doi{10.1038/s41586-021-03382-w}.

\bibitem{Bailey2015}
Bailey JE, Nagayama T, Loisel GP, Rochau GA, Blancard C, Colgan J, et~al.
\newblock A higher-than-predicted measurement of iron opacity at solar interior
  temperatures.
\newblock Nature. 2015 Jan;517(7532):56--59.
\newblock \doi{10.1038/nature14048}.

\bibitem{Liu2019}
Liu SF, Hori Y, M{\"u}ller S, Zheng X, Helled R, Lin D, et~al.
\newblock The formation of Jupiter's diluted core by a giant impact.
\newblock Nature. 2019 Aug;572(7769):355--357.
\newblock \doi{10.1038/s41586-019-1470-2}.

\bibitem{Brygoo2021}
Brygoo S, Loubeyre P, Millot M, Rygg JR, Celliers PM, Eggert JH, et~al.
\newblock Evidence of hydrogen-helium immiscibility at Jupiter-interior
  conditions.
\newblock Nature. 2021 May;593(7860):517--521.
\newblock \doi{10.1038/s41586-021-03516-0}.

\bibitem{Kraus_Science_2022}
Kraus RG, Hemley RJ, Ali SJ, Belof JL, Benedict LX, Bernier J, et~al.
\newblock Measuring the melting curve of iron at super-Earth core conditions.
\newblock Science. 2022;375(6577):202--205.
\newblock \doi{10.1126/science.abm1472}.

\bibitem{Kritcher_Nature_2020}
Kritcher AL, Swift DC, D{\"o}ppner T, Bachmann B, Benedict LX, Collins GW,
  et~al.
\newblock A measurement of the equation of state of carbon envelopes of white
  dwarfs.
\newblock Nature. 2020 Aug;584(7819):51--54.
\newblock \doi{10.1038/s41586-020-2535-y}.

\bibitem{becker}
Becker A, Lorenzen W, Fortney JJ, Nettelmann N, Sch\"ottler M, Redmer R.
\newblock Ab initio equations of state for hydrogen (H-REOS.3) and helium
  (He-REOS.3) and their implications for the interior of brown dwarfs.
\newblock Astrophys J Suppl Ser. 2014;215:21.
\newblock \doi{10.1088/0067-0049/215/2/21}.

\bibitem{Kraus2016}
Kraus D, Ravasio A, Gauthier M, Gericke DO, Vorberger J, Frydrych S, et~al.
\newblock Nanosecond formation of diamond and lonsdaleite by shock compression
  of graphite.
\newblock Nature Communications. 2016 Mar;7(1):10970.
\newblock \doi{10.1038/ncomms10970}.

\bibitem{Kraus2017}
Kraus D, Vorberger J, Pak A, Hartley NJ, Fletcher LB, Frydrych S, et~al.
\newblock Formation of diamonds in laser-compressed hydrocarbons at planetary
  interior conditions.
\newblock Nature Astronomy. 2017 Sep;1(9):606--611.
\newblock \doi{10.1038/s41550-017-0219-9}.

\bibitem{hu_ICF}
Hu SX, Militzer B, Goncharov VN, Skupsky S.
\newblock First-principles equation-of-state table of deuterium for inertial
  confinement fusion applications.
\newblock Phys Rev B. 2011 Dec;84:224109.
\newblock \doi{10.1103/PhysRevB.84.224109}.

\bibitem{Betti2016}
Betti R, Hurricane OA.
\newblock Inertial-confinement fusion with lasers.
\newblock Nature Physics. 2016 May;12(5):435--448.
\newblock \doi{10.1038/nphys3736}.

\bibitem{Zylstra2022}
Zylstra AB, Hurricane OA, Callahan DA, Kritcher AL, Ralph JE, Robey HF, et~al.
\newblock Burning plasma achieved in inertial fusion.
\newblock Nature. 2022 Jan;601(7894):542--548.
\newblock \doi{10.1038/s41586-021-04281-w}.

\bibitem{Tschentscher_2017}
Tschentscher T, Bressler C, Grünert J, Madsen A, Mancuso AP, Meyer M, et~al.
\newblock Photon Beam Transport and Scientific Instruments at the European
  XFEL.
\newblock Applied Sciences. 2017;7(6).
\newblock \doi{10.3390/app7060592}.

\bibitem{SACLA_2011}
Pile D.
\newblock First light from SACLA.
\newblock Nature Photonics. 2011 Aug;5(8):456--457.
\newblock \doi{10.1038/nphoton.2011.178}.

\bibitem{LCLS_2016}
Bostedt C, Boutet S, Fritz DM, Huang Z, Lee HJ, Lemke HT, et~al.
\newblock Linac Coherent Light Source: The first five years.
\newblock Rev Mod Phys. 2016 Mar;88:015007.
\newblock \doi{10.1103/RevModPhys.88.015007}.

\bibitem{OMEGA}
Soures JM, McCrory RL, Verdon CP, Babushkin A, Bahr RE, Boehly TR, et~al.
\newblock {Direct‐drive laser‐fusion experiments with the OMEGA, 60‐beam,
  $>$40 kJ, ultraviolet laser system}.
\newblock Physics of Plasmas. 1996 05;3(5):2108--2112.
\newblock \doi{10.1063/1.871662}.

\bibitem{sinars2020review}
Sinars D, Sweeney M, Alexander C, Ampleford D, Ao T, Apruzese J, et~al.
\newblock Review of pulsed power-driven high energy density physics research on
  Z at Sandia.
\newblock Physics of Plasmas. 2020;27(7).
\newblock \doi{10.1063/5.0007476}.

\bibitem{Moses_NIF}
Moses EI, Boyd RN, Remington BA, Keane CJ, Al-Ayat R.
\newblock The National Ignition Facility: Ushering in a new age for high energy
  density science.
\newblock Physics of Plasmas. 2009;16(4):041006.
\newblock \doi{10.1063/1.3116505}.

\bibitem{MacDonald_POP_2023}
MacDonald MJ, Di~Stefano CA, Döppner T, Fletcher LB, Flippo KA, Kalantar D,
  et~al.
\newblock {The colliding planar shocks platform to study warm dense matter at
  the National Ignition Facility}.
\newblock Physics of Plasmas. 2023 06;30(6).
\newblock 062701. \doi{10.1063/5.0146624}.

\bibitem{siegfried_review}
Glenzer SH, Redmer R.
\newblock X-ray Thomson scattering in high energy density plasmas.
\newblock Rev Mod Phys. 2009;81:1625.
\newblock \doi{10.1103/RevModPhys.81.1625}.

\bibitem{Gregori_PRE_2003}
Gregori G, Glenzer SH, Rozmus W, Lee RW, Landen OL.
\newblock Theoretical model of x-ray scattering as a dense matter probe.
\newblock Phys Rev E. 2003 Feb;67:026412.
\newblock \doi{10.1103/PhysRevE.67.026412}.

\bibitem{PhysRevE.94.053211}
Harbour L, Dharma-wardana MWC, Klug DD, Lewis LJ.
\newblock Pair potentials for warm dense matter and their application to x-ray
  Thomson scattering in aluminum and beryllium.
\newblock Phys Rev E. 2016 Nov;94:053211.
\newblock \doi{10.1103/PhysRevE.94.053211}.

\bibitem{GarciaSaiz2008}
Garc{\'i}a~Saiz E, Gregori G, Gericke DO, Vorberger J, Barbrel B, Clarke RJ,
  et~al.
\newblock Probing warm dense lithium by inelastic X-ray scattering.
\newblock Nature Physics. 2008 Dec;4(12):940--944.
\newblock \doi{10.1038/nphys1103}.

\bibitem{Tilo_Nature_2023}
D{\"o}ppner T, Bethkenhagen M, Kraus D, Neumayer P, Chapman DA, Bachmann B,
  et~al.
\newblock Observing the onset of pressure-driven K-shell delocalization.
\newblock Nature. 2023 May;\doi{10.1038/s41586-023-05996-8}.

\bibitem{Dornheim_T_2022}
Dornheim T, B{\"o}hme M, Kraus D, D{\"o}ppner T, Preston TR, Moldabekov ZA,
  et~al.
\newblock Accurate temperature diagnostics for matter under extreme conditions.
\newblock Nature Communications. 2022 Dec;13(1):7911.
\newblock \doi{10.1038/s41467-022-35578-7}.

\bibitem{dornheim2023xray}
Dornheim T, Döppner T, Baczewski AD, Tolias P, Böhme MP, Moldabekov ZA,
  et~al.
\newblock X-ray Thomson scattering absolute intensity from the f-sum rule in
  the imaginary-time domain.
\newblock arXiv. 2023;{\href{https://arxiv.org/abs/2305.15305}{{2305.15305}}}.
  {[physics.plasm-ph]}.

\bibitem{Regan_PRL_2012}
Regan SP, Falk K, Gregori G, Radha PB, Hu SX, Boehly TR, et~al.
\newblock Inelastic X-Ray Scattering from Shocked Liquid Deuterium.
\newblock Phys Rev Lett. 2012 Dec;109:265003.
\newblock \doi{10.1103/PhysRevLett.109.265003}.

\bibitem{Falk_PRE_2013}
Falk K, Regan SP, Vorberger J, Crowley BJB, Glenzer SH, Hu SX, et~al.
\newblock Comparison between x-ray scattering and velocity-interferometry
  measurements from shocked liquid deuterium.
\newblock Phys Rev E. 2013 Apr;87:043112.
\newblock \doi{10.1103/PhysRevE.87.043112}.

\bibitem{Falk_PRL_2014}
Falk K, Gamboa EJ, Kagan G, Montgomery DS, Srinivasan B, Tzeferacos P, et~al.
\newblock Equation of State Measurements of Warm Dense Carbon Using
  Laser-Driven Shock and Release Technique.
\newblock Phys Rev Lett. 2014 Apr;112:155003.
\newblock \doi{10.1103/PhysRevLett.112.155003}.

\bibitem{PhysRevLett.107.115004}
Caillabet L, Canaud B, Salin G, Mazevet S, Loubeyre P.
\newblock Change in Inertial Confinement Fusion Implosions upon Using an Ab
  Initio Multiphase DT Equation of State.
\newblock Phys Rev Lett. 2011 Sep;107:115004.
\newblock \doi{10.1103/PhysRevLett.107.115004}.

\bibitem{Hurricane_Nature_2014}
Hurricane OA, Callahan DA, Casey DT, Celliers PM, Cerjan C, Dewald EL, et~al.
\newblock Fuel gain exceeding unity in an inertially confined fusion implosion.
\newblock Nature. 2014 Feb;506(7488):343--348.
\newblock \doi{10.1038/nature13008}.

\bibitem{Chihara_1987}
Chihara J.
\newblock Difference in X-ray scattering between metallic and non-metallic
  liquids due to conduction electrons.
\newblock Journal of Physics F: Metal Physics. 1987 feb;17(2):295--304.
\newblock \doi{10.1088/0305-4608/17/2/002}.

\bibitem{Baczewski_PRL_2016}
Baczewski AD, Shulenburger L, Desjarlais MP, Hansen SB, Magyar RJ.
\newblock X-ray Thomson Scattering in Warm Dense Matter without the Chihara
  Decomposition.
\newblock Phys Rev Lett. 2016 Mar;116:115004.
\newblock \doi{10.1103/PhysRevLett.116.115004}.

\bibitem{new_POP}
Bonitz M, Dornheim T, Moldabekov ZA, Zhang S, Hamann P, Kählert H, et~al.
\newblock Ab initio simulation of warm dense matter.
\newblock Physics of Plasmas. 2020;27(4):042710.
\newblock \doi{10.1063/1.5143225}.

\bibitem{Dornheim_review}
Dornheim T, Moldabekov ZA, Ramakrishna K, Tolias P, Baczewski AD, Kraus D,
  et~al.
\newblock {Electronic density response of warm dense matter}.
\newblock Physics of Plasmas. 2023 03;30(3).
\newblock 032705. \doi{10.1063/5.0138955}.

\bibitem{kraus_xrts}
Kraus D, Bachmann B, Barbrel B, Falcone RW, Fletcher LB, Frydrych S, et~al.
\newblock Characterizing the ionization potential depression in dense carbon
  plasmas with high-precision spectrally resolved x-ray scattering.
\newblock Plasma Physics and Controlled Fusion. 2018 nov;61(1):014015.
\newblock \doi{10.1088/1361-6587/aadd6c}.

\bibitem{Dornheim_T2_2022}
Dornheim T, Böhme MP, Chapman DA, Kraus D, Preston TR, Moldabekov ZA, et~al.
\newblock {Imaginary-time correlation function thermometry: A new,
  high-accuracy and model-free temperature analysis technique for x-ray Thomson
  scattering data}.
\newblock Physics of Plasmas. 2023 04;30(4).
\newblock 042707. \doi{10.1063/5.0139560}.

\bibitem{quantum_theory}
Giuliani G, Vignale G.
\newblock Quantum Theory of the Electron Liquid.
\newblock Cambridge: Cambridge University Press; 2008.

\bibitem{DOPPNER2009182}
Döppner T, Landen OL, Lee HJ, Neumayer P, Regan SP, Glenzer SH.
\newblock Temperature measurement through detailed balance in x-ray Thomson
  scattering.
\newblock High Energy Density Physics. 2009;5(3):182--186.
\newblock \doi{10.1016/j.hedp.2009.05.012}.

\bibitem{raymond1976radiative}
Raymond JC, Cox DP, Smith BW.
\newblock Radiative cooling of a low-density plasma.
\newblock The Astrophysical Journal. 1976;204:290--292.
\newblock \doi{10.1086/154170}.

\bibitem{van1954photon}
Van~Roosbroeck W, Shockley W.
\newblock Photon-radiative recombination of electrons and holes in germanium.
\newblock Physical Review. 1954;94(6):1558.
\newblock \doi{10.1103/PhysRev.94.1558}.

\bibitem{thorsheim1987laser}
Thorsheim H, Weiner J, Julienne PS.
\newblock Laser-induced photoassociation of ultracold sodium atoms.
\newblock Physical review letters. 1987;58(23):2420.
\newblock \doi{10.1103/PhysRevLett.58.2420}.

\bibitem{MacDonald_POP_2022}
MacDonald MJ, Saunders AM, Bachmann B, Bethkenhagen M, Divol L, Doyle MD,
  et~al.
\newblock Demonstration of a laser-driven, narrow spectral bandwidth x-ray
  source for collective x-ray scattering experiments.
\newblock Physics of Plasmas. 2021;28(3):032708.
\newblock \doi{10.1063/5.0030958}.

\bibitem{CHUNG20053}
Chung HK, Chen MH, Morgan WL, Ralchenko Y, Lee RW.
\newblock FLYCHK: Generalized population kinetics and spectral model for rapid
  spectroscopic analysis for all elements.
\newblock High Energy Density Physics. 2005;1(1):3--12.
\newblock \doi{10.1016/j.hedp.2005.07.001}.

\bibitem{Voigt_POP_2021}
Voigt K, Zhang M, Ramakrishna K, Amouretti A, Appel K, Brambrink E, et~al.
\newblock {Demonstration of an x-ray Raman spectroscopy setup to study warm
  dense carbon at the high energy density instrument of European XFEL}.
\newblock Physics of Plasmas. 2021;28(8):082701.
\newblock \doi{10.1063/5.0048150}.

\bibitem{Militzer_PRE_2021}
Militzer B, Gonz\'alez-Cataldo F, Zhang S, Driver KP, Soubiran Fmc.
\newblock First-principles equation of state database for warm dense matter
  computation.
\newblock Phys Rev E. 2021 Jan;103:013203.
\newblock \doi{10.1103/PhysRevE.103.013203}.

\bibitem{Bohme_PRL_2022}
B\"ohme M, Moldabekov ZA, Vorberger J, Dornheim T.
\newblock Static Electronic Density Response of Warm Dense Hydrogen: Ab Initio
  Path Integral Monte Carlo Simulations.
\newblock Phys Rev Lett. 2022 Aug;129:066402.
\newblock \doi{10.1103/PhysRevLett.129.066402}.

\end{thebibliography}

\section*{Acknowledgments}
We gratefully acknowledge helpful comments by Marius Millot.

This work was partly funded by the Center for Advanced Systems Understanding (CASUS) which is financed by Germany's Federal Ministry of Education and Research (BMBF) and by the Saxon Ministry for Science, Culture and Tourism (SMWK) with tax funds on the basis of the budget approved by the Saxon State Parliament. This work has received funding from the European Research Council (ERC) under the European Union’s Horizon 2022 research and innovation programme
(Grant agreement No. 101076233, "PREXTREME"). 
Sandia National Laboratories is a multimission laboratory managed and operated by National Technology and Engineering Solutions of Sandia, LLC, a wholly owned subsidiary of Honeywell International Inc., for the U.S. Department of Energy's National Nuclear Security Administration under Contract No. DE-NA0003525.
The work of Ti.~D., M.~J.~M, and F.R.G.~was performed under the auspices of the U.S. Department of Energy by Lawrence Livermore National Laboratory under Contract No. DE-AC52-07NA27344.

\section*{Author Contributions Statement}

M.B., J.V., and To.D.~developed the idea, carried out the analysis, and wrote substantial parts of the 
manuscript. L.B.F.~and Ti.D.~substantially contributed to the analysis and to writing the manuscript. D.K., A.D.B., T.R.P., M.J.M., F.R.G., and Zh.A.M. contributed to the analysis and to writing the manuscript.

\section*{Competing Interests Statement}

The authors declare no competing interests.

\end{document}